\documentclass[12pt]{book}

\usepackage[dvips]{graphicx,color}
\usepackage{makeidx,view}

\makeauthorindex
\makeindex

\def\cf{{cf.~}}

\def\lsim{\raise0.3ex\hbox{$<$}\kern-0.75em{\lower0.65ex\hbox{$\sim$}}} 
\def\gsim{\raise0.3ex\hbox{$>$}\kern-0.75em{\lower0.65ex\hbox{$\sim$}}} 
\def\sc1{\raise2.1ex\hbox{\tiny $r\!\!=\!\!4$}\kern-0.95em{\hbox{$=$}}}

\def\cm3{~{\rm cm^{-3}}}

\def\hinv{$h^{-1}$}

\def\ltsima{$\; \buildrel < \over \sim \;$}
\def\simlt{\lower.5ex\hbox{\ltsima}}
\def\gtsima{$\; \buildrel > \over \sim \;$}
\def\simgt{\lower.5ex\hbox{\gtsima}}

\def\app{{\rm Astropart.~Phys.\ }}

\def\nat{{\rm Nature \ (London) \ }}

\def\prp{{\rm Phys.~Rep.\ }}

\def\sc{{\rm Science\ }}

\def\cpc{{\rm Comp.\ Phys.\ Comm.\ }}

\newcommand{\gr}{$\gamma$-ray }
\newcommand{\epm}{e$^\pm$ }

\newcommand{\pnd}{$\pi^0$-decay }

\newcommand{\ic}{IC }

\begin{document}

\BookTitle{\itshape The Universe Viewed in Gamma-Rays}
\CopyRight{\copyright 2002 by Universal Academy Press, Inc.}
\pagenumbering{arabic}

\chapter{
Investigating Galaxy Clusters through $\gamma$-ray Emission
}

\author{%
Francesc Miniati
\\
{\it Max-Planck-Institut f\"ur Astrophysik,
     Karl-Schwarzschild-Str. 1, 85740, Garching, Germany}}

%
\AuthorContents{F.\ Miniati} 

\AuthorIndex{Miniati}{F.}

\section*{Abstract}

We address the role of $\gamma$-ray astronomy in
the investigation of nonthermal 
processes in the large scale structure of the universe.
Based on EGRET upper limits on nearby galaxy clusters (GCs)
we constrain the acceleration efficiency of CR electrons 
at intergalactic shocks to $\leq 1 \% $ than the shock 
ram pressure. That implies a contribution to the 
cosmic $\gamma$-ray background from intergalactic shocks
of order 25 \% of the measured level.
We model spatial and spectral properties of 
nonthermal \gr emission due to shock accelerated 
cosmic-rays (CRs) in GCs and {\it emphasize} the
importance of imaging capability of upcoming
\gr facilities for a correct interpretation of 
the observational results.
GC observations at this photon 
energy will help us understand the
origin of the radio emitting particles,
the possible level of CR pressure
and the strength of magnetic fields
in intracluster environment and possibly will 
allow us detect the accretion shocks.

\section{Introduction}

Cosmic shocks emerge during structure formation in the
universe, due to gravitationally driven supersonic gas 
infall onto collapsing objects. 
Astrophysical shocks are collisionless and
capable of accelerating cosmic-rays (CRs) via first 
order Fermi mechanism [2].
CR electrons suffer severe energy losses 
dominated, in typical intra-cluster environment, 
by inverse Compton (IC) for energies above 
$\sim 150$ MeV and Coulomb losses below that.
Low energy, sub-relativistic CR protons also have
short lifetimes due to Coulomb losses. 
However, for relativistic
protons the dominant energy loss mechanism 
up to the energy threshold for photo-pair production
is p-p inelastic collisions with a timescale longer than 
a Hubble time. Thus, once accelerated CR protons
accumulate within large scale structures (LSSs) confined there
by $\mu$G strong turbulent magnetic fields [22].

It is of interest to investigate CR acceleration at
LSS shocks for several reasons. 
Firstly, GCs exhibit non-thermal radiation.
This mainly consists of diffuse radio emission extending on 
Mpc scales. It is thought to be synchrotron radiation 
but there is no consensus as to the origin of the emitting 
electrons. 
Also, excess of radiation with respect 
to thermal emission has been reported at hard X-ray 
[7] and, more controversially, 
extreme UV photon energies
[10,4].
Secondly, if shocks acceleration operates efficiently,
the proton component could bear a significant fraction of the 
total gas pressure, affecting the dynamics of cosmic structures 
[18].
Finally, CR electrons accelerated at inter-galactic shocks
could contribute to the cosmic \gr background
(CGB) [11].
This contribution addresses the role of \gr observations
for investigating CR acceleration at LSS shocks.
We describe the method in \S \ref{num.sec}, the results in 
\ref{cgb.sec}-\ref{coma.sec} 
and conclude in \S \ref{disc.sec}


\section{Numerical Simulation} \label{num.sec}

We a perform numerical simulation that models
simultaneously the formation of the LSS
and the evolution of three CR species: 
namely {\it shock accelerated} protons and (primary) electrons
and secondary \epm generated in p-p inelastic collisions of the 
simulated CR protons and the thermal gas.
For the LSS we adopt a canonical, flat $\Lambda$CDM model
with total mass density $\Omega_m=0.3$,
vacuum energy density $\Omega_\Lambda= 0.7$,
Hubble constant
$H_0=67 $ km s$^{-1}$ Mpc$^{-1}$,
 baryonic mass density, $\Omega_b=0.04$, and spectral index
and cluster normalization for initial density perturbations, 
$n_s=1$ and $\sigma_8=0.9$ respectively (see ref. [14] for 
full details).

The CR dynamics, includes: injection at shocks, acceleration,
energy losses and spatial transport. It is computed 
numerically through the code COSMOCR [13,14].
According to the adopted thermal leakage injection scheme, 
a fraction about $10^{-4}$ of the protons crossing the shock 
is injected as CRs. 
For shocks with Mach number, $M= 4 -10$,
responsible for most of the heating of the intergalactic gas [14], 
this roughly corresponds to converting 30\% of the shock 
ram pressure into CR proton pressure.
Further, we assume that the ratio of accelerated primary CR 
electrons to protons at relativistic energies is 
given by a parameter $R_{e/p}\sim 10^{-2}$ [12,1].
In accord to the test particle limit of diffusive shock acceleration
theory [2], the accelerated particles are distributed in 
momentum as a power-law with slope $q=4/(1-M^{-2})$. 
All relevant loss mechanisms are accounted for [13,14]. 
We follow CR protons in the momentum range between $0.1$ GeV/c and
$10^6$ GeV/c, and CR electrons and \epm between 15 MeV/c and 20 TeV/c.

\section{Contribution to the Cosmic Gamma-ray Background}
\label{cgb.sec}
\begin{figure}
\begin{center}
\rule{0cm}{0mm}
\includegraphics[height=7cm]{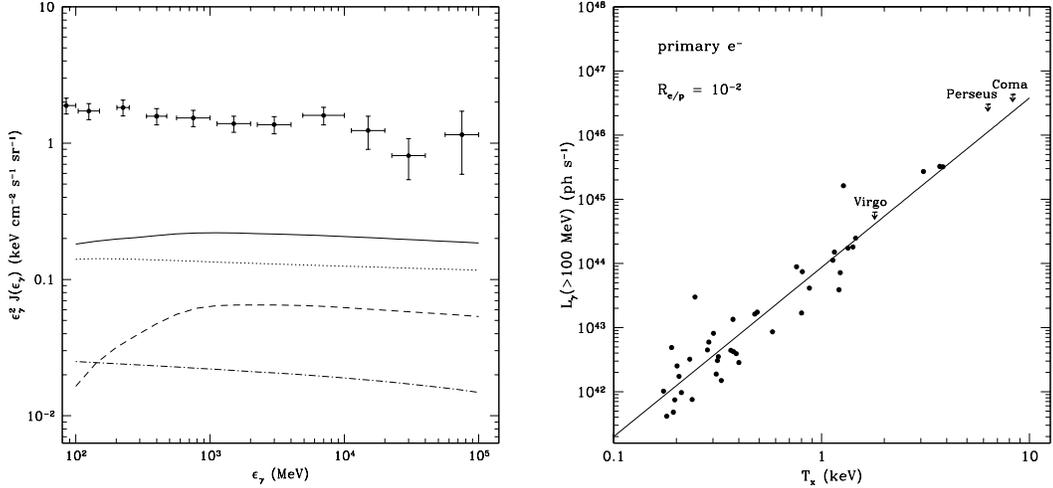}
\end{center}
\caption{{\it Left:} 
$\gamma$-ray background flux produced by 
cosmological CRs and EGRET data with their error-bars 
(solid dots) [21].
{\it Right:} The IC $\gamma$-ray photon luminosity 
above 100 MeV from individual clusters as a function 
of the cluster X-ray core temperature, $T_x$. 
EGRET experimental upper limits are from Reimer [19].
\label{gamma.fig} }
\end{figure}

Fig. \ref{gamma.fig} (left) shows the 
contribution to the CGB from CRs accelerated at 
LSS shocks [14]. It is computed as 
\begin{equation}
\varepsilon^2 J(\varepsilon)
= \varepsilon \, \frac{c}{4\pi H_0} \; \int_{0}^{z_{max}}
\frac{e^{-\tau_{\gamma\gamma}}}{[\Omega_m (1+z)^3 + \Omega_\Lambda]^{1/2}} \;
\frac{j[\varepsilon (1+z),z]}{(1+z)^4} \; dz
\end{equation}
where $\varepsilon$ is the photon energy,
$j(\varepsilon ,z)$ is the computational-box-averaged
spectral emissivity in units `photons cm$^{-3}$ s$^{-1}$'
at red-shift, $z$, $\tau_{\gamma\gamma}$
is an attenuation factor due to photo-pair creation,
$\gamma\gamma \rightarrow e^\pm$, $c$ is the speed of 
light and $z_{max}$ an upper limit of integration. 
EGRET observational data (solid dots) [21] are also shown for
comparison.
We consider the following emission processes: IC emission 
of CR electrons scattering off cosmic microwave background 
photons (dot line), decay of neutral pions 
($\pi^0 \rightarrow \gamma\gamma$)
produced in p-p inelastic collisions (dash line) and
\ic emission from secondary \epm (dot-dash line). 
In fig. \ref{gamma.fig} the total flux (solid line) 
corresponds roughly to a constant value at the level of 
0.2 keV cm$^{-2}$ s$^{-1}$ sr$^{-1}$ throughout the spectrum. 
It is dominated by \ic emission from primary 
electrons. Fractions of 
order 30\% and 10 \% are produced by $\pi^0$-decay 
and \ic emission from secondary \epm respectively.

All three components produce the 
same flat spectrum, consistent with the observations.
The result is a reflection of the fact that the CRs
distributions were generated in strong shocks [14].
However, the computed flux is only $\sim$ 15 \% of the 
observed CGB. It is difficult to imagine a higher 
contribution from \pnd and IC emission from 
\epm. In fact, if more CR protons were produced
at shocks, CR-induced shock modifications would actually
reduce the population of \gr emitting protons (and \epm).
On the other hand, the fraction, $\eta$, of shock ram 
pressure converted into 
CR electrons, can be constrained by
comparing  the simulated clusters' \gr photon luminosity
above 100 MeV to the upper limits set by the EGRET 
[20,19] for nearby GCs.
This is done in fig. \ref{gamma.fig} (right panel).
The simulation data are best-fit by the curve (solid line):
\begin{equation} \label{gamfit.eq}
{\rm L}_\gamma (>100~ {\rm MeV})= 8.7\times 10^{43}  
\left(\frac{\eta}{4\times 10^{-3}}\right)
\; \left(\frac{\rm T_x}{\rm keV}\right)^{2.6} \; {\rm ph ~s}^{-1}.
\end{equation}
According to our calculations [14], 
the EGRET upper limits require that 
$\eta \leq 0.8 \%$. This implies an upper limit on the computed 
\gr flux of about 0.35 keV cm$^{-2}$ s$^{-1}$ sr$^{-1}$
or a fraction of order $\sim 25$ \%  of the CGB.  

\section{Synthetic emission maps and spectrum for a Coma-like prototype}
\label{coma.sec}

In this section we explore the spectral and spatial properties
of \gr radiation between 10 keV and 10 TeV due to 
shock accelerated CRs in GCs. 
As non-thermal bremsstrahlung turns out
unimportant [15], we consider the following emission 
processes: \pnd and IC emission from both primary CR electrons 
and secondary \epm. We first compute the emission spectrum 
for the largest virialized object in the computational box
which has an X-ray core temperature $T_x\simeq 4$ keV. 
The various emission components are then renormalized to the
case of a Coma-like cluster. 
In particular, IC emission 
from electrons is rescaled according to equation \ref{gamfit.eq}.
The total number of \epm is rescaled assuming
that these particles
are responsible the for synchrotron emission of Coma radio halo
\footnote{This assumption is not necessary but it is assumed 
for simplicity}. For 
the purpose we took a radio flux at 1.4 GHz $S_{1.4GHz} = 640$ mJy 
[5] and a two values for the volume average magnetic field 
$\langle B \rangle$, namely 1 and 3 $\mu$G [9].
\begin{figure}
\begin{center}
\includegraphics[height=5.3cm]{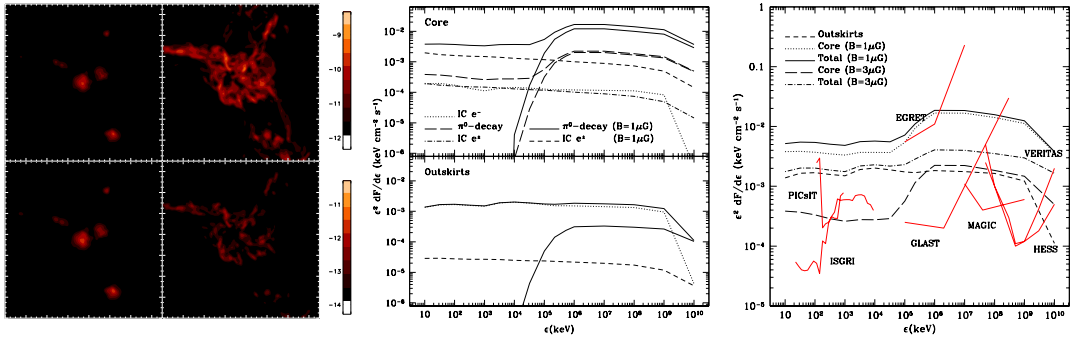}
\end{center}
\caption{{\it Left:} 
Synthetic map of the integrated photon flux 
in units ``ph cm$^{-2}$ s$^{-1}$ arcmin$^{-2}$''. 
Each panel measures 15 \hinv Mpc on a side.
{\it Center:} 
Synthetic spectra from 10 keV up to 10 TeV extracted from a
core and an outskirts region.
{\it Right:} 
Comparison of synthetic spectra with nominal sensitivity limits 
of future \gr observatories (thick-solid lines). 
Significance and observing time are respectively
 3$\sigma$ and $10^6$ s
for INTEGRAL-IBIS imagers (ISGRI and PICsIT), 
5$\sigma$ and one year of all sky survey for
EGRET and GLAST and 5$\sigma$ and 50 hour exposure
for the Cherenkov telescopes. 
\label{figc4.fig}}
\end{figure}

In the left quadrant of fig. \ref{figc4.fig} we present 
synthetic maps of the integrated photon flux above 100 keV 
(top panels) and 100 MeV (bottom panels).
They show that the emission from \pnd (bottom left) and 
\epm (top left) is confined to the cluster core
where it creates a diffuse 
halo which rapidly fades with distance from the center.
In fact, \epm and $\pi^0$ are produced at the 
highest rate in the densest regions where both the parent CR ions 
and target nuclei are most numerous. 
On the other hand, because of severe energy losses, 
\gr emitting primary electrons are only found 
in the vicinity of strong shocks where they are accelerated. 
The emission (right panels) is extended and with a rich
morphology reflecting the complex ``web'' of accretion 
shocks surrounding GCs [17]. 
The corresponding 
synthetic spectra extracted from a core (top; with a $0.5^o$ radius) 
and an outskirts region (bottom; a ring with inner and outer 
radii of $0.5^o$ and $1.5^o$ respectively) are shown
in the central quadrant of fig. \ref{figc4.fig}.
They shows that the emitted radiation 
in the outskirts region is strongly 
dominated by \ic emission from primary primary e$^-$.
Conversely in the core region $\pi^0$-decay (solid thin line)
dominates at high photon energy ($>$ 100 MeV) (top panel).
Below $\sim$ 100 MeV the relative contribution from 
primary and secondary \epm depends on the strength of the 
magnetic field.
Finally, in the right quadrant the predicted
radiation spectra in various spatial regions are compared
to sensitivity limits\footnote{These are for point sources and 
will be worse for extended sources considered here.}
of planned \gr observatories. 

\section{Conclusions} \label{disc.sec}
Clusters observations at $\gamma$-ray energy 
will provide important information about GCs.
First, the extended emission is produced at the location 
of accretion shocks which, unlike weak merging 
shocks, have yet to be observed {\it directly}. 
Thus, detection and imaging of \ic emission from primary 
electrons would provide an opportunity for their direct 
observation.
The flux about 100 keV together 
with radio measurements, will allow an estimate of 
the intracluster magnetic field.
It is important, however, to separate out the contribution
from the CR electrons accelerated at the outlining shocks.
In fact, because the magnetic field strength is expected to drop at the
cluster outskirts, these electrons likely generate only a weak radio 
emission. Nonetheless, they produce a strong \ic flux which can easily
dominate the total soft \gr emission (\cf fig. \ref{figc4.fig}).
Measuring the \gr flux at and above 100 MeV will allow us to
confirm or rule out secondary \epm models for radio 
halos in GC [6,3,16]
and to estimate the non-thermal CR pressure 
there [18]. In this perspective several authors 
estimated for nearby clusters 
the \gr flux from $\pi^0$-decay. However, radiation flux from
\ic emission can be comparable [15]. Therefore, 
for a correct interpretation of the measurements the 
contributions from these two
components will need to be separated as outlined here.

Measuring the spectrum of IC \gr from primary
CR electrons will also be instrumental for addressing 
their contribution to the CGB 
[11,14,8], the details of their acceleration mechanism 
and the physical conditions of the shocks.

\section{Acknowledgments}
This work is partially supported by the European Community
under the contract HPRN-CT2000-00126 RG29185 
(The Physics of the Intergalactic Medium).
The computational work was carried out at the
the Rechenzentrum in Garching.



\vspace{1pc}

\re
1. Allen, G.~E., Petre, R., \& Gotthelf, E.~V. 2001, ApJ, 558, 739
\re
2. Blandford, R.~D. \& Eichler, D. 1987, \prp, 154, 1
\re
3. Blasi, P. \& Colafrancesco, S. 1999, \app, 12, 169
\re
4. Bowyer, S., Berghoefer, T.~W., \& Korpela, E. 1999, ApJ, 526, 592
\re
5. Deiss, B.~M., Reich, W., Lesch, H., \& Wielebinski, R. 1997, A\&A, 321, 55
\re
6. Dolag, K. \& En{\ss}lin, T. 2000, A\&A, 362, 151
\re
7. Fusco-Femiano, R., et al. 1999, ApJ, 513, L21
\re
8. Keshet, U., et~al. 2002, ApJ, in press (astro-ph/0202318)
\re
9. Kim, K.-T., et~al. 1990, ApJ, 355, 29
\re
10. Lieu, R., et al. 1996, \sc, 274, 1335
\re
11. Loeb, A. \& Waxman, E. 2000, \nat, 405, 156
\re
12. M\"uller, D. \& et~al. 1995, in Int. Cosmic Ray Conference, Vol.~3, Rome, 13
\re
13. Miniati, F. 2001, \cpc, 141, 17
\re
14. Miniati, F. 2002, MNRAS, 337, 199
\re
15. Miniati, F. 2002, MNRAS, submitted
\re
16. Miniati, F., Jones, T.~W., Kang, H., \& Ryu, D. 2001, ApJ, 562, 233
\re
17. Miniati, F., et~al. 2000, ApJ, 542, 608
\re
18. Miniati, F., Ryu, D., Kang, H., \& Jones, T.~W. 2001, ApJ, 559,
\re
19. Reimer, O. 1999, in ICRC, Vol.~4, ed. D.~Kieda, et al., Salt Lake City, 89
\re
20. Sreekumar, et al. 1996, ApJ, 464, 628
\re
21. Sreekumar, et~al. 1998, ApJ, 494, 523
\re
22. V\"{o}lk, H.~J., Aharonian, F.~A., \& Breitschwerdt, D. 1996, Sp.Sci.Rev., 75, 279

\endofpaper
\end{document}